\documentclass[12pt,epsf]{article}
\usepackage{amsmath}
\usepackage{amssymb}
\begin{document}
\def\a{\alpha}
\def\b{\beta}
\def\c{\varepsilon}
\def\d{\delta}
\def\e{\epsilon}
\def\f{\phi}
\def\g{\gamma}
\def\h{\theta}
\def\k{\kappa}
\def\l{\lambda}
\def\m{\mu}
\def\n{\nu}
\def\p{\psi}
\def\q{\partial}
\def\r{\rho}
\def\s{\sigma}
\def\t{\tau}
\def\u{\upsilon}
\def\v{\varphi}
\def\w{\omega}
\def\x{\xi}
\def\y{\eta}
\def\z{\zeta}
\def\D{{\mit \Delta}}
\def\G{\Gamma}
\def\H{\Theta}
\def\L{\Lambda}
\def\F{\Phi}
\def\P{\Psi}

\def\S{\Sigma}

\def\o{\over}
\def\beq{\begin{eqnarray}}
\def\eeq{\end{eqnarray}}
\newcommand{\gsim}{ \mathop{}_{\textstyle \sim}^{\textstyle >} }
\newcommand{\lsim}{ \mathop{}_{\textstyle \sim}^{\textstyle <} }
\newcommand{\vev}[1]{ \left\langle {#1} \right\rangle }
\newcommand{\bra}[1]{ \langle {#1} | }
\newcommand{\ket}[1]{ | {#1} \rangle }
\newcommand{\EV}{ {\rm eV} }
\newcommand{\KEV}{ {\rm keV} }
\newcommand{\MEV}{ {\rm MeV} }
\newcommand{\GEV}{ {\rm GeV} }
\newcommand{\TEV}{ {\rm TeV} }
\def\diag{\mathop{\rm diag}\nolimits}
\def\Spin{\mathop{\rm Spin}}
\def\SO{\mathop{\rm SO}}
\def\O{\mathop{\rm O}}
\def\SU{\mathop{\rm SU}}
\def\U{\mathop{\rm U}}
\def\Sp{\mathop{\rm Sp}}
\def\SL{\mathop{\rm SL}}
\def\tr{\mathop{\rm tr}}

\baselineskip 0.7cm

\begin{titlepage}

\begin{flushright}
UT-KOMABA/07-7
\end{flushright}

\vskip 1.35cm
\begin{center}
{\large \bf
D-dualized D-brane}
\vskip 1.2cm
Yu Nakayama${}^{1}$

\vskip 0.4cm

${}^1${\it Institute of Physics, University of Tokyo,\\
      Komaba, Meguro-ku, Tokyo 153-8902, Japan}

\vskip 2.5cm

\abstract{We further investigate the dimensional duality (D-duality) proposed in arXiv:0705.0550 by mainly focusing on the properties of D-branes in this background. We derive the world-sheet correspondence of static D-branes, and discuss the fate of non-static D-branes from the world-sheet viewpoint. The quantum  string production with or without D-branes is also studied from the time-like Liouville theory. We find that the closed string production from the background is much larger than that from D-branes decaying into nothing.}
\end{center}
\end{titlepage}

\setcounter{page}{2}

\section{Introduction}
One of the central issues in theoretical physics --- especially quantum gravity including the string theory, is to understand the initial singularity of the universe. Observations of the current universe, together with the singularity theorem, make it almost inevitable that our universe began with the singularity at the level of Einstein's general relativity.

Recently the use of the (winding) tachyon condensation to remove such a classical singularity was proposed \cite{McGreevy:2005ci} (see also \cite{Hikida:2005ec,She:2005qq,Nakayama:2006gt,McInnes:2006uz,Hellerman:2006nx,Brandenberger:2007xu} for related works). Closed string tachyon condensation often gives rise to dimensionality change in the target space, which is particular to string/M-theory. The perturbative duality discussed in \cite{Green:2007tr} is one concrete example of such a scenario.\footnote{More general arguments for the dimensional change in supercritical string theories can be found in \cite{Hellerman:2004zm,Hellerman:2004qa,Hellerman:2006ff,Hellerman:2007fc}}

This dimensional duality is formulated within the perturbative string theory, and it relates the string theory compactified on a genus $h$ Riemann surface $M_h$ and the string theory compactified on a $2h$-dimensional torus $T^{2h}$. The duality is intrinsically time-dependent. This is because on one hand, the genus $h$ Riemann surface (for $h>1$) has negative curvature and does not solve the Einstein equation by itself and we need extra time-dependence. On the other hand, the compactification on $T^{2h}$ is supercritical in general, so we have to introduce the time-like linear dilaton in order to compensate the excess of the central charge. The world-sheet renormalization group argument\footnote{See e.g. \cite{Freedman:2005wx} for a recent argument on the relationship between the time-dependence and the renormalization group flow} shows that the infrared limit (future) corresponds to the genus $h$ Riemann surface, and the ultraviolet limit (past) corresponds to $T^{2h}$ as we will review in section \ref{section2}.

In this paper, we investigate D-duality by mainly focusing on the properties of D-branes. A crucial clue to discover D-duality pairs in \cite{Green:2007tr} was the winding number conservation \cite{McGreevy:2006hk}. However, in string theory, we have many other higher dimensional (non-perturbative) objects such as D-branes. To understand  D-duality better, a precise map between these D-branes under D-duality should be of importance. For example, one may pose a question: since the structure of the higher homotopy group (or homology group) is different between the D-dual pairs (i.e. $\pi_n(M_h) \neq \pi_n(T^{2h}$) in general), what happens to the charges of higher dimensional D-branes? It should be contrasted against the winding number conservation of the fundamental string.

We would also like to ask the stability of this time-dependent duality under the quantum production of closed strings. The dimensional duality involves closed string tachyon condensation, so such time-dependence is expected to show a large back reaction due to the quantum closed string production. In this paper, we revisit this problem from the time-like Liouville theory approach. This approach gives a specific analytic continuation scheme to evaluate such closed string production as Bogoliubov transformation.

The organization of this paper is as follows. In section \ref{section2}, we introduce D-duality as proposed in \cite{Green:2007tr}. We review their world-sheet proof and we generalize their results in  a broader cosmological context. Then, we discuss the quantum closed string production from the time-like Liouville theory. In section \ref{section3}, we investigate D-dualized D-branes. We first give a world-sheet proof of D-duality with boundaries. The fate of the D-brane which cannot be D-dualized from the world-sheet theory will also be studied by using the boundary renormalization group flow and the time-like boundary Liouville theory. In section \ref{section4}, we give further discussions and conclude the whole paper.

\section{Bulk D-duality}\label{section2}
D-duality is the duality between the string theory compactified on a  genus $h$ Riemann surface and the string theory compactified on a $2h$-dimensional torus. The world-sheet derivation of the duality was given in \cite{Green:2007tr}, and we begin with the review of their derivation.\footnote{The path integral derivation  reviewed here is similar to the one for T-duality \cite{Buscher:1987sk,Buscher:1987qj}}
\subsection{setup}
Let us consider the following bulk action \cite{Green:2007tr}
\begin{align}
S &= \int d^2\sigma \gamma_{ab} \biggl\{ (\partial Y^a-A^a)(\partial \bar{Y}^b -\bar{A}^b) - (X^a\bar{F}^b + F^a\bar{X}^b)  \cr
 &+ \left. [G^a(\bar{X}^b-\int^{\bar{Z}}\bar{\omega}^b)+(X^a-\int^{Z}\omega^a)\bar{G}^b]  + \frac{1}{e^2}G^a\bar{G}^b \right\} \ , \label{bulka}
\end{align} 
where we have two kinds of $U(1)$ (complex) gauge fields: $F^a = \partial A^a$, $G^a= \partial B^a$. Here $\omega^a$ ($a=1,\cdots,h$) are independent holomorphic 1-forms on a given genus $h$ Riemann surface $M_h$, and $\gamma_{ab}$ is a positive constant matrix. The action has a gauge invariance under
\begin{align}
Y^a &\to Y^a + \Lambda^a \cr
A^a &\to A^a + \partial \Lambda^a \ .  \label{gauge}
\end{align}
We rewrite the action in two different ways to give a world-sheet proof of D-duality.

We first integrate out $X^a$ fields as Lagrange multipliers, which demands $F^a = G^a$. Now by using the gauge freedom, we set $Y^a=A^a-B^a = 0$. The action reads
\begin{eqnarray}
S' = \int d^2\sigma \gamma_{ab} \left\{ A^a\bar{A}^b - F^a\int^{\bar{Z}}\bar{\omega}^b - \int^Z \omega^a \bar{F}^b + \frac{1}{e^2} F^a\bar{F}^b \right\}
\end{eqnarray}
After integrating out $A^a$, we obtain
\begin{eqnarray}
S' = \int d^2\sigma  \omega^a \partial Z \gamma_{ab} \bar{\omega}^b \partial \bar{Z} 
\end{eqnarray}
in the infrared  limit by neglecting higher derivative terms. The effective action describes a non-linear sigma model on a genus $h$ Riemann surface.

Alternatively, one can take $Y^a=0$ gauge first, and integrate out $A^a$. This results in
\begin{align}
S'' &= \int d^2\sigma \gamma_{ab} \left\{ \partial X^a\partial X^b + G^a(\bar{X}^b-\int^{\bar{Z}}\bar{\omega}^b) + (X^a- \int^Z \omega^a)\bar{G}^b + \frac{1}{e^2}G^a\bar{G}^b \right\} \cr
 &= \int d^2\sigma \gamma_{ab} \partial X^a \partial X^b - U(X) \ ,
\end{align}
where the 1-loop effective potential \cite{Coleman:1976uz,Witten:1993yc} is given by
\begin{eqnarray}
U(X) = e^2 (X^a - \int^Z \omega^a)\gamma_{ab}(\bar{X}^b-\int^{\bar{Z}}\bar{\omega}^b) \ .
\end{eqnarray}
If we neglect the potential in the ultraviolet limit, we obtain the target space $T^{2h}$ as the Jacobian torus of the genus $h$ Riemann surface.\footnote{The Jacobian torus is defined by the identification $X^a \sim X^a + \Omega^a$, where $\Omega^a = \oint_{\gamma_b} \omega^a$ is a period matrix.}

Here, we have a comment on general properties of the duality. In the derivation above, explicit form of the map $\int^Z \omega^a$ does not play any important role at all.\footnote{The author would like thank Y.~Kikukawa for a related discussion.} Indeed, one can easily generalize the argument in the following way. Consider arbitrary holomorphic embedding of the complex manifold $Z$ into $X$: $f^a(Z^\alpha) \in X^a$. One can now repeat the same argument above by replacing the Abel-Jacobi map $\int^Z \omega^a$ by $f^a(Z^\alpha)$. 

On one hand, we obtain the effective action describing a non-linear sigma model on $Z$ with the action
\begin{eqnarray}
S = \int d^2\sigma \frac{\partial f^a}{\partial Z^\alpha}\frac{\partial \bar{f}^b}{\partial \bar{Z}^\beta} \gamma_{ab} \partial Z^\alpha \partial \bar{Z}^\beta \ .
\end{eqnarray}
On the other hand, we have a dual description as a linear sigma model on $X$ with the potential
\begin{eqnarray}
S_{dual} = \int d^2\sigma \gamma_{ab}\partial X^a\partial\bar{X}^b - U(X) \ ,
\end{eqnarray}
where
\begin{eqnarray}
U(X) = e^2 (X^a - f^a(Z)) \gamma_{ab} (\bar{X}^b-\bar{f}^b(Z)) \ .
\end{eqnarray}

As a low energy effective field theory description, the duality between the two is simply equivalence between the non-linear sigma model and the linear sigma model with the potential that vanishes on the embedding map. As a string duality, one should note that the infrared limit (later time) is realized as the non-linear sigma model with the target space $Z$ and the ultraviolet limit (earlier time) is realized by the linear sigma model with the target space $X$. In the very early universe, one can neglect the world-sheet potential, and the dimensionality of the target space is enhanced. From the target space viewpoint, the tachyon potential induces the dimensional transition from $X$ to $Z$.

The realization of the ultraviolet configuration sensitively depends on the tachyon potential. In other words, by cleverly choosing the form of the tachyon potential, one can create desirable universes as a consequence of the closed string tachyon condensation. 

This gives a new insight into the avoidance of the space-time singularity of the universe from the string theory. If our space-like section of the universe is created as a consequence of the tachyon condensation from the supercritical string theory, an apparent singularity at an early universe from the viewpoint of the observer on $Z$ is removed by dualizing the theory to a higher dimensional manifold $X$ (with dilaton gradient). One can say that the tachyon condensation created the universe from the supercritical linear dilaton theory.\footnote{Still, depending on the sign of the linear dilaton charge, the string theory might be strongly coupled in the earlier universe. One interesting way to resolve the singularity is introduce the light-like linear dilaton as studied in \cite{Craps:2005wd,Ishino:2005ru,Ishino:2006nx}}

\subsection{Large quantum back reaction}
At the early times, the process of D-duality can be approximated by the time-like linear dilaton theory perturbed by the growing tachyon potential. It is interesting to see the back reaction of the tachyon condensation from the closed string production. If one further assumes the mild space-dependence of the tachyon condensation, or if one focuses on the dominant constant mode of the tachyon condensation under the Fourier decomposition, it will be described by the time-like Liouville theory.

The action of the original Liouville theory (see \cite{Nakayama:2004vk} for a review) is given by
\begin{eqnarray}
S_{Liouville} = \int d^2\sigma \left(\partial \phi \bar{\partial}\phi + \frac{(b+b^{-1})R\phi}{4} +\mu e^{2b\phi} \right) \ , 
\end{eqnarray}
where $R$ is the world-sheet curvature.
We perform the analytic continuation $\phi \to iT$, $b\to-i\beta$ to obtain
\begin{eqnarray}
S_{t-Liouville} = \int d^2\sigma \left(-\partial T\bar{\partial}T + \frac{(\beta - \beta^{-1})RT}{4} + \mu e^{2\beta T} \right) \ .
\end{eqnarray}
We take $\beta <1$, as expected from the Seiberg bound in the space-like Liouville theory: $b<1$, so that the setup is precisely what is proposed in \cite{Green:2007tr}. We will come back to the case $\beta>1$ later.\footnote{Note that $\beta \to 0$ is the classical limit, where the mini-superspace approximation is reliable. The mini-superspace approximation of the model was studied in \cite{Aharony:2006ra}.}

The closed string production of the model is described by the Bogoliubov coefficient appearing in the two-point function of the time-like Liouville theory given by
\begin{align}
P_\omega = |\gamma_{\omega}|^2 &= \left|-(\pi\mu\gamma(-\beta^2))^{-2i\omega/\beta} \frac{\Gamma(1+2i\omega/\beta)\Gamma(1-2i\omega \beta)}{\Gamma(1-2i\omega/\beta)\Gamma(1+2i\omega\beta)}\right|^2 \cr
 &= e^{-4\pi\omega/\beta} \label{parc}
\end{align}
with a specific analytic continuation scheme that agrees with the mini-superspace result.\footnote{We will not discuss a subtlety associated with three-point functions. See \cite{Schomerus:2003vv,Fredenhagen:2003ut,Zamolodchikov:2005fy,McElgin:2007ak} for recent discussions.}

The particle creation is exponentially suppressed as the energy increases, but in string theory, we have exponentially growing number of string modes as a function of the energy or mass, which results in Hagedorn divergence. In the time-like Liouville theory at hand, by recalling that the central charge is given by $c = 1 - 6(\beta-\beta^{-1})^2$, we obtain the density of states\footnote{We assume that there is no mismatch between the central charge and the effective central charge except for the time-like linear dilaton part.} 
\begin{eqnarray}
\rho(M) \sim e^{2\pi M(\beta+\beta^{-1})} \ ,
\end{eqnarray}
where $M$ is the mass of the string states. Therefore, the total particle creation is
\begin{eqnarray}
N \simeq \int dM |\gamma_{\omega=\frac{M}{2}}|^2 \rho (M) \sim \int dM e^{-{2\pi M}{\beta^{-1}}} e^{2\pi M(\beta+\beta^{-1})} \ .
\end{eqnarray}
Since ${2\pi M}{\beta}^{-1}<{2\pi M(\beta+\beta^{-1})}$ for any $\beta \ (<1)$, the string creation shows exponential divergence (see also \cite{Strominger:2003fn} for a particular case with $\beta =1$). 

Our result shows that the tachyon condensation process associated with D-duality accompanies a huge amount of the closed string production. The effect of such string production possibly with a cosmological implication seems interesting and worth further studying.\footnote{A similar large back reaction was observed in another asymptotically time-like dilaton theory in \cite{Toumbas:2004fe}.} 

Finally as mentioned before, let us briefly study another possible choice of the analytic continuation: $\beta >1$, where we do not see any initial singularity as $T\to -\infty$. In this case, the quantum particle creation \eqref{parc} depends sensitively on the value of $\beta$
\begin{align}
|\gamma_\omega|^2 &= 1  \ \  \mathrm{when}  \ \ ( 2n-1 <\beta^2 <2n) \cr
|\gamma_{\omega}|^2 &=  e^{-4\pi\omega/\beta} \ \ \mathrm{when}  \ \ ( 2n <\beta^2 <2n+1)
\end{align}
for an integer $n$.
This peculiar behavior of the quantum particle creation, which deviates from the mini-superspace result, was also observed in the time-like Sine-Liouville theory \cite{Nakayama:2006gt}. In either case, the closed string production is larger than the mini-superspace regime discussed above, and the back reaction is enormous. Furthermore, the string coupling becomes stronger in later time, so the cosmological scenario for $\beta>1$ is even more complicated.

\section{D-dualized D-brane}\label{section3}
\subsection{Boundary correspondence}
To understand the properties of the D-dual geometry better, let us discuss D-branes in the D-dual geometry.\footnote{As in the closed string case, the following argument is in close resemblance to the one for the path integral proof of the open string T-duality \cite{Alvarez:1996up,Dorn:1996an}.} In this subsection, we study static D-branes, and in the next subsection, we discuss time-dependent D-branes.

We begin with the D0-brane in the D-dualized geometry. To describe the D0-brane, we add the boundary interaction
\begin{eqnarray}
 S_{bound} = \int ds \gamma_{ab} \left\{ c^a(X_{\perp}) (-\partial_n \bar{Y}^b +\bar{A}^b_n) \dot{\sigma}^n + \bar{c}^a(X_{\perp}) (-\partial_n Y^b + A^b_n) \dot{\sigma}^n \right\} \ . \label{bound}
\end{eqnarray}
to the original action \eqref{bulka} now defined on the world-sheet with boundaries.
Note that this boundary interaction is gauge invariant under \eqref{gauge}.
For technical reasons, we restrict ourselves, for a time being, to the case when $c^{a}(X_{\perp})  = \int^Z{\omega}^a$ for a particular $Z$ (i.e. $c^{a}$ is on the Riemann surface). Here $X_{\perp}$ denotes transverse directions, and the dependence of $c^a$ on $X_{\perp}$ should be determined by the (quantum) boundary conformal invariance. As is the case with the bulk theory, we rewrite the boundary interaction in two different ways to obtain the D-branes in the D-dual geometry.

We first integrate out $X^a$. By a gauge choice, we can set $Y^a=A^a-B^a=0$. Then, integrating over $A^a$ introduces a surface term, which, after combining it with the original contribution from \eqref{bound}, gives
\begin{eqnarray}
S'_{bound} = \int ds \gamma_{ab} \left\{ c^a(X_{\perp})\bar{A}^b_{n} -\int^Z \omega^a \bar{A}^b_{n} + \bar{c}^a(X_{\perp}){A}^b_{n} -\int^{\bar{Z}} \bar{\omega}^a {A}^b_{n} \right\}\dot{\sigma}^n  \ .
\end{eqnarray}
The integration over $A^a$ is ultralocal (in the large $e^2$ limit), so we obtain the delta functional constraint
\begin{eqnarray}
\delta ( c^a (X_{\perp}) - \int^z \omega^a)|_{bound}  \ , \label{posi1}
\end{eqnarray}
which is the Dirichlet boundary condition for a D0-brane localized on the genus $h$ Riemann surface.

Alternatively, let us take the $Y^a=0$ gauge and integrate out $A^a$ first. The induced surface term together with the original boundary action \eqref{bound} yields the boundary action
\begin{eqnarray}
S''_{bound} = \int ds \gamma_{ab} \left\{c^a(X_{\perp}) \bar{A}^b_{n} - X^a\bar{A}^b_{n} + \bar{c}^a(X_{\perp}){A}^b_{n} - \bar{X}^a{A}^b_{n} \right\} \dot{\sigma}^n \ .
\end{eqnarray}
Again, the integration over $A^a$ is ultralocal, so we obtain the delta functional constraint
\begin{eqnarray}
\delta ( c^a (X_{\perp}) - X^a)|_{bound}  \ , \label{posi2}
\end{eqnarray}
which is the Dirichlet boundary condition for a D0-brane localized on the Jacobian torus $T^{2n}$. The location of the D0-brane is determined by the Abel-Jacobi map: $ X^a = c^a (X_{\perp}) = \int^Z \omega^a$.

Let us move on to the D2-brane. We consider the boundary interaction
\begin{eqnarray}
 S_{bound} = \int ds \gamma_{ab} \left\{ \int^Z{\omega}^a (-\partial_n \bar{Y}^b +\bar{A}^b_n) \dot{\sigma}^n + \int^{\bar{Z}}{\bar{\omega}}^a (-\partial_n Y^b + A^b_n) \dot{\sigma}^n \right\} \ . \label{bound2}
\end{eqnarray}

We first integrate out $X^a$. By a gauge choice, we can set $Y^a=A^a-B^a=0$. Then, integrating over $A^a$ introduces a surface term that cancels with \eqref{bound2}. Thus, there is no boundary interaction at the end, which forces the Neumann boundary condition for the D2-brane on the genus $h$ Riemann surface. 

Alternatively, let us take the $Y^a=0$ gauge and integrate out $A^a$ first. The induced surface term yields the boundary action
\begin{eqnarray}
S''_{bound} = \int ds \gamma_{ab} \left\{\int^Z{\omega}^a \bar{A}^b_{n} - X^a\bar{A}^b_{n} + \int^{\bar{Z}}\bar{\omega}^a{A}^b_{n} - \bar{X}^a{A}^b_{n} \right\} \dot{\sigma}^n \ .
\end{eqnarray}
The subsequent integration over $A^a$ gives the constraint
\begin{eqnarray}
\delta ( \int^Z{\omega}^a - X^a)|_{bound}  \ .
\end{eqnarray}
This delta functional means that the D2-brane on the Jacobian torus $T^{2h}$ should be localized along the image of the Abel-Jacobi map from the genus $h$ Riemann surface.

In either case, one can introduce the gauge field on the D2-brane by adding
\begin{eqnarray}
\int ds \gamma_{ab} \left\{ \mathcal{A}^a(X_{\perp}) \bar{\omega}^b \partial_n \bar{Z} +\bar{\mathcal{A}}^a(X_{\perp}) {\omega}^b \partial_n Z \right\} \dot{\sigma}^n
\end{eqnarray}
to the boundary interaction \eqref{bound2}. The moduli space of the gauge field is given by the dual torus $(T^{2h})^*$ realized by the Wilson line on the D2-brane \cite{Green:2007tr}.

Let us finally discuss the D1-brane. We parametrize the D1-brane on the genus $h$ Riemann surface by the one-parameter embedding $c^a(t) \in \int^z \omega $ with a parameter $t$. The boundary interaction would be
\begin{eqnarray}
\int ds \gamma_{ab} \left\{ c^a(t) (-\partial_n \bar{Y}^b +\bar{A}^b_n) \dot{\sigma}^n + \bar{c}^a(t) (-\partial_n Y^b + A^b_n) \dot{\sigma}^n \right\} \ . \label{bound3}
\end{eqnarray}
In the path integral, we also integrate over $t$. 

We first integrate out $X^a$. By a gauge choice, we can set $Y^a=A^a-B^a=0$. Then, integrating over $A^a$ gives a delta functional constraint
\begin{eqnarray}
\int dt \delta(c^a(t)-\int^Z\omega^a)|_{bound}  \ .
\end{eqnarray}
This condition gives the D1-brane on the genus $h$ Riemann surface.

Alternatively, let us take the $Y^a=0$ gauge and integrate out $A^a$ first.
 It gives the boundary constraint
\begin{eqnarray}
\int dt \delta(c^a(t)-X^a)|_{bound} \ .
\end{eqnarray}
This condition determines the shape of the D1-brane on the Jacobian torus $T^{2h}$ as the image of the D1-brane under the Abel-Jacobi map. One could also introduce the gauge field on the D1-brane.

So far, we have restricted ourselves to the case when the D-brane is localized on the Abel-Jacobi map from the genus $h$ Riemann surface. With this restriction,  we have seen that D$p$-branes on the Riemann surface correspond to D$p$-branes on the Jacobian torus, which are given by the image of the Abel-Jacobi map. However, this restriction seems unnecessary from the viewpoint of the world-sheet theory on $T^{2h}$. For example, we could have studied the D0-brane outside of the image of the Abel-Jacobi map by considering $c^a(X_{\perp})$ in \eqref{bound} outside of the image of the Abel-Jacobi map. Then, at first sight, \eqref{posi2} gives the right position for such D0-brane. However, it is inconsistent with \eqref{posi1} and the partition function must vanish. 

Actually, this can be seen from the original action  \eqref{bulka} with the boundary. After a partial differentiation, we have a boundary interaction
\begin{eqnarray}
S_{bound} = \int ds \gamma_{ab} \left\{ B_n^a(\bar{X}^b-\int^{\bar{Z}}\bar{\omega}^b)+  \bar{B}_n^a({X}^b -\int^{Z}{\omega}^b) \right\} \dot{\sigma}^n \ . \label{part}
\end{eqnarray}
The integration over ${B^a}$ is ultralocal (in the large coupling limit: $e^2\to\infty$), so it gives the boundary condition
\begin{eqnarray}
\delta(X^a-\int^Z \omega^a)|_{bound} \ .
\end{eqnarray}
Thus, the D0-brane outside of the Abel-Jacobi map is not allowed in this setup. The discussion with other D$p$-brane is completely in parallel. The physical implication of this result will be further discussed in the next subsection.

\subsection{Fate of other D-branes}
As we discussed, the support of the D-branes should be on the genus $h$ Riemann surface (or its Abel-Jacobi map onto the torus) in the low-energy effective field theory. However, in the far ultraviolet regime, one could prescribe the D-brane not localized on the Abel-Jacobi map. This is because in $e^2 \to 0$ limit, ultralocality of $B^a$ integral  in \eqref{part} is violated. Thus, one may expect the existence of the D-branes extending along the total $2h$-dimensional torus $T^{2h}$ not restricted to the Abel-Jacobi map. Of course, the construction of such a D-brane is always possible if we begin with the world-sheet theory with the flat torus target space with no tachyon condensation.

In this subsection, we study the fate of such D-branes from two different perspectives. The first approach is to use the world-sheet (boundary) renormalization group flow interpreted as time evolution. The second approach is to use the analytic continuation of the D-branes in Liouville field theory. Both analyses show that under the time evolution, the D-branes are forced to be on the position where D-duality is effective.

{\bf Renormalization group flow analysis}

One way to trace the time-dependence of the D-brane under the closed string tachyon condensation is to use the correspondence between the time evolution and the boundary renormalization group flow. Under the general deformation of the boundary CFT
\begin{eqnarray}
S = S_0 + \sum_i \lambda_i \int d^2\sigma \phi_i(\sigma) + \sum_a\mu_a \int ds \psi_a(s) \ ,
\end{eqnarray}
we have a perturbative renormalization group equation \cite{Fredenhagen:2006dn,Green:2006ku}
\begin{align}
\dot{\lambda}_i &= (2-\Delta_{\phi_i})\lambda_i + C_{ijk}\lambda_j\lambda_k + \cdots \cr\dot{\mu}_a &= (1-\Delta_{\psi_a})\mu_a + B_{ai}\lambda_i + D_{abc}\mu_b\mu_c + \cdots \ ,
\end{align}
where $\Delta$ denotes the conformal dimension of the operator and $C_{ijk}$, $B_{ai}$, $D_{abc}$ denote bulk, bulk-boundary and  boundary structure constants respectively.

When the D0-brane on $T^{2h}$ is located outside of the image under the Abel-Jacobi map, $\mu \psi(s)$ is given by $\gamma_{ab}(\delta c^{a} \bar{A}^b_n + \delta \bar{c}^aA^b_n) \dot{\sigma}^n$. The tachyon condensation is given by the closed string perturbation $\lambda\phi(\sigma) = U(X)$ at the classical level. It is clear when $U(c^a)=0$, the classical OPE vanishes and the closed string tachyon condensation does not affect the position of the D0-brane at this order of the perturbation theory.\footnote{At the quantum level, this is not always true due to the curvature of the target space. Solutions of the effective DBI action might not be static.} In contrast, when $U(c^a) \neq 0$, the bulk-boundary OPE $B_{ia}$ does not vanish and the D-brane feels force attracted toward the zero-potential locus, where world-sheet D-duality is feasible.

{\bf Boundary time-like Liouville analysis}

As in the closed string background, we can approximate the fate of the D-brane in the tachyon condensation region by the analytic continuation of the boundary Liouville theory. The disappearing brane in the future would correspond to the time-like analogue of the FZZT brane \cite{Fateev:2000ik,Teschner:2000md,Gutperle:2003xf}.

The FZZT brane has the boundary action
\begin{eqnarray}
\int ds \left( \frac{(b+b^{-1})K\phi}{2\pi} + \mu_B e^{b\phi} \right) \ ,
\end{eqnarray}
where $K$ is the boundary curvature.
The same analytic continuation discussed in section 2 gives the disappearing D-brane into the future in the time-like Liouville theory
\begin{eqnarray}
\int ds \left( \frac{(\beta-\beta^{-1})KT}{2\pi} + \mu_B e^{\beta T} \right) \ .
\end{eqnarray}
A part of the reason why we have a disappearing D-brane in this setup is the open string tachyon condensation, but the physical quantities such as the radiation rate at the world-sheet 1-loop approximation does not depend on the boundary cosmological constant $\mu_B$, so we may argue that the following results are also applicable when $\mu_B = 0$ where we expect that the disappearance of the D-brane is due to the closed string tachyon condensation.

To discuss the closed string radiation from such a disappearing D-brane, we compute the absolute square of the disk one-point function.\footnote{Alternatively, one can study the imaginary part of the cylinder amplitude. Formally these two computations are related by the optical theorem, but in practice the computation involves a subtle analytic continuation as thoroughly discussed in \cite{Nakayama:2006qm}} The disk one-point function of the FZZT brane is given by
\begin{eqnarray}
U(p) = \frac{2}{b} (\pi\mu\gamma(b^2))^{-\frac{ip}{b}} \Gamma(1+2ipb)\Gamma(2ipb^{-1}) \cos(2\pi sp) \ ,
\end{eqnarray}
where $\cosh^2\pi bs = \frac{\mu_B^2}{\mu} \sin\pi b^2$ \cite{Fateev:2000ik}.

After the analytic continuation, one can see that the radiation rate behaves as
\begin{eqnarray}
P_\omega = |U(\omega)|^2 \sim e^{-2\pi\omega(\beta +\beta^{-1})} \ 
\end{eqnarray}
for large $\omega$. The total radiation rate can be estimated as
\begin{eqnarray}
N \simeq \int dM P_{\omega=\frac{M}{2}} \sqrt{\rho(M)} \sim \int dM e^{-\pi M(\beta +\beta^{-1})} e^{+\pi M(\beta +\beta^{-1})} \ ,
\end{eqnarray}
which shows a power-like behavior irrespective of the value of $\beta$. This shows that the radiation rate from the disappearing brane is potentially diverging, but this simply comes from the total energy of the D-brane that also diverges (as $1/g_s$) in the string perturbation theory. The argument for the universal behavior of the decaying D-brane --- independence of $\beta$ in our case, can be found in \cite{Nakayama:2005pk,Nakayama:2006qm,Nakayama:2007sb}. The closed string radiation is not so large compared with the direct back reaction due to the closed string tachyon condensation, and we conclude that D-brane decays into nothing without disturbing the D-dualized geometry.

\section{Discussion and Conclusion}\label{section4}
In this paper, we have investigated the properties of D-duality. D-duality is a  novel mechanism to engineer the universe by using the closed string tachyon condensation from the time-like linear dilaton theory. As we discussed in section 2, the closed string tachyon condensation in general produces a large back reaction. 

The effects of such a back reaction, especially on the remaining universe are of great significance in the context of cosmology. The initial value problem of the universe is crucial to understand the vacuum selection of the string compactification and the inflation of the universe. If the universe is made out of such closed string tachyon condensation, the universe will be dominated by the closed string radiation. Time evolution of such a universe would be worth studying and the spectrum density discussed in this paper would be important.

We have also discussed D-branes in the D-dual geometry. Unlike the winding number for the fundamental string, the higher topological charges of the D-branes are not generally conserved under D-duality. In such cases, D-branes fall apart in pieces into nothing during the closed string tachyon condensation process. The energy stored with the D-brane will be radiated away as closed string emission. In contrast, we have also found that there exist other kinds of D-branes invariant under D-duality. These D-branes are static under the closed string tachyon condensation.

In this paper, we have approximated the dynamics of the D-branes by neglecting the spatial dependence of the tachyon condensation. To understand the inhomogeneous decay of the D-branes, we need to go beyond this approximation. An analytic continuation of the boundary states in Sine-Liouville theory would be helpful although they are only available for a specific parameter region (such as $\mathcal{N}=2$ Liouville point) at this moment.\footnote{In the mini-superspace approximation, the spatial dependence of the tachyon condensation tends to reduce the string creation. It would be very interesting to study this effect from the exact CFT approach. The author would like to thank E.~Silverstein for the inspiring discussions and for sharing the idea.}  

\section*{Acknowledgements}
The author thanks Y.~Kikukawa and E.~Silverstein for the discussion on a generalization of D-duality, and for the discussion on the string creation under the space dependent tachyon condensation, respectively.
He also acknowledges the Japan Society for the Promotion of Science for financial
support.

\end{document}